\documentclass[twocolumn,english,aps,prb,twocolum,superscriptaddress,natbib,bibnotes,amsmath,amssymb,floatfix,groupedaddress,footinbib]{revtex4-2}

\usepackage[colorlinks=true,citecolor=blue,linkcolor=magenta]{hyperref}

\usepackage[markup=nocolor, authormarkupposition=left]{changes} 

\usepackage{soul}
\usepackage[utf8]{inputenc}
\usepackage[english]{babel}
\usepackage{amsmath,amsfonts,amssymb}
\usepackage[T1]{fontenc}
\usepackage{url}

\usepackage{amsmath}
\usepackage{siunitx}
\usepackage[version=4]{mhchem}
\usepackage{amsfonts}
\usepackage{amssymb}

\usepackage{epstopdf}
\usepackage{graphicx}
\graphicspath{{./Figures/}}

\usepackage{changes}

\begin{document}
\title{Generalized angle-orbital-angular-momentum Talbot effect and modulo mode sorting}

\author{Jianqi Hu$^{1,\dagger,*}$, Matias Eriksson$^{2,\dagger}$, Sylvain Gigan$^{1}$ \& Robert Fickler$^{2}$}
\affiliation{
$^1$Laboratoire Kastler Brossel\char`,{} École Normale Supérieure - Paris Sciences et Lettres (PSL) Research University\char`,{} Sorbonne Université\char`,{} Centre National de la Recherche Scientifique (CNRS)\char`,{} UMR 8552\char`,{} Collège de France\char`,{} 24 rue Lhomond\char`,{} 75005 Paris\char`,{} France.\\
$^2$Physics Unit\char`,{} Photonics Laboratory\char`,{} Tampere University\char`,{} 33720 Tampere\char`,{} Finland.\\
}

\maketitle

\noindent\textbf{\noindent
The Talbot effect describes periodic revivals of field patterns and is ubiquitous across wave systems. In optics, it is mostly known for its manifestations in space and time, but is also observed in the wavevector and frequency spectra owing to the Fourier duality. Recently, the Talbot self-imaging has been shown separately in the azimuthal angle and orbital angular momentum (OAM) domains. Here, we unveil the missing link between them and demonstrate the generalized angle-OAM Talbot effect. Versatile transformations of petal fields and OAM spectra are experimentally showcased, based on the synergy of angular Talbot phase modulation and light propagation in a ring-core fiber.
Moreover, the generalized self-imaging concept leads to new realizations in mode sorting, which separate OAM modes in a modulo manner, theoretically free from any crosstalk within the congruence classes of OAM modes. We design and experimentally construct various mode sorters with excellent performance, and show the unconventional behavior of Talbot-based sorters where neighboring OAM modes can be mapped to positions far apart. Besides its fundamental interest, our work finds applications in OAM-based information processing, and implies that the physical phenomena in time-frequency and angle-OAM domains are broadly connected as well as their processing techniques may be borrowed interchangeably. }

\section*{Introduction} 

\noindent{The} Talbot effect is a ubiquitous wave self-imaging phenomenon that occurs in diverse wave systems. It is firstly observed by Henry Fox Talbot in the spatial degree of freedom \cite{talbot1836lxxvi}, when light wave traverses periodic structures and experiences Fresnel diffraction \cite{rayleigh1881x}. The spatial self-imaging of a monochromatic optical field delineates the renowned Talbot carpet \cite{berry2001quantum,case2009realization,wen2013talbot}, where identical or multiplied optical patterns are periodically revived in the near-field propagation. Less widely known is the Fourier analogue of the spatial Talbot effect that appears in the far-field, manifested by the self-imaging of transverse wavevectors of field patterns in the angular spectrum \cite{azana2014angular}. Besides spatial representations, the Talbot effect also takes place in the time-frequency domain arising from the space-time duality \cite{kolner1994space}. The temporal Talbot effect describes linear propagation of periodic optical pulse trains through dispersive elements \cite{azana2001temporal}, e.g., chirped fiber Bragg gratings \cite{longhi200040} or a spool of optical fiber \cite{meloni2005250}, transforming the pulse train into an identical or a repetition-rate multiplied replica of the input. From the spectral perspective, such processes can be comprehended as assigning specific quadratic phases, i.e. Talbot phases \cite{cortes2016generality,fernandez2017structure}, to the constituent frequency comb lines of the pulse train. 
In contrast to the temporal Talbot effect, the spectral self-imaging is realized by Talbot phase modulation of a periodic optical pulse train in the time domain \cite{azana2005spectral}, typically via either electro-optic modulation \cite{caraquitena2011spectral} or nonlinear optical processes \cite{lei2015observation}. While preserving the overall spectral envelope, the spectral Talbot effect redistributes energy in the optical spectrum and generates new frequency components in between the initial comb lines. 
The concepts of temporal and spectral Talbot self-imaging have recently been unified, leading to the establishment of the generalized Talbot effect \cite{cortes2016generality,romero2019arbitrary}. In essence, the combined time-frequency Talbot effect allows for arbitrary control of the pulse repetition rate as well as the corresponding comb free spectral range. This technique enables passive amplification and denoising of optical waveforms \cite{maram2014noiseless,crockett2022optical} and is in general useful for optical communications, microwave photonics, and spectroscopy. 

Orbital angular momentum (OAM) is another degree of freedom associated with light, which describes light beams with helical phase fronts \cite{yao2011orbital}. 
It is arguably one of the most important mode bases in structured light \cite{rubinsztein2016roadmap,forbes2021structured} and has led to fundamental studies \cite{fickler2016quantum} and diverse applications ranging from communications \cite{wang2012terabit}, optical tweezers \cite{padgett2011tweezers}, 
sensing \cite{lavery2013detection}, to quantum information processing \cite{erhard2018twisted}. The Fourier-related domain of OAM is the transverse azimuthal angle\cite{Yao:06}. Notably, an OAM mode resembles a frequency mode in many ways.
A superposition of OAM modes is analogous to an optical frequency comb, while the azimuthal angle 
is analogous to time with one repetition period bounded between $0$ to $2\pi$. Such a connection has been identified, and has been utilized for the generation, manipulation, and measurement of OAM states, leveraging techniques from the time-frequency domain \cite{xie2017spatial, yang2019manipulation, lin2023single}. Additionally, both self-imaging phenomena in the azimuthal angle \cite{niemeier1985self,baranova1998talbot,hautakorpi2006modal,samadian2016cylindrical,eriksson2021talbot} and OAM spectrum \cite{lin2021spectral} have been observed, akin to the temporal and spectral Talbot effects. 
The angular Talbot effect, i.e., self-imaging in the azimuthal angle, arises when an azimuthally localized field is launched into a ring-core fiber (RCF) \cite{eriksson2021talbot}. During propagation, the localized input pattern reemerges periodically at integer and fractional Talbot lengths.
The underlying mechanism is rooted in the Talbot phases acquired by the OAM spectrum of the input pattern \cite{hu2018talbot}. 
Conversely, the OAM Talbot effect renders self-imaging of modes in the OAM spectrum, by passing evenly-spaced angular petals through a phase mask that azimuthally modulates the field with Talbot phases \cite{lin2021spectral}.
Nevertheless, although both angular and OAM Talbot effects have been reported, the nexus between them and the potential of the combined effect remains unexplored. 

In this paper, we predict and experimentally demonstrate the generalized angle-OAM Talbot effect, showcasing the unique capability to flexibly tailor the number of angular petals and OAM mode spacings. Moreover, the combined angle-OAM Talbot effect leads to new functionalities in mode sorting, which is instrumental in OAM-based systems yet challenging to achieve high-performance experimentally. 
We conceive a scalable OAM sorting scheme based on the combination of Talbot phase masks \cite{lin2021spectral} and RCF \cite{eriksson2021talbot}, designed following the generalized Talbot effect. This scheme sorts OAM modes in a modulo manner, i.e., OAM orders incongruent modulo an arbitrary predefined integer $q$ are steered to non-overlapping angular sectors, while orders congruent modulo $q$ are mapped to the same angular sectors. We experimentally demonstrate a few examples of such sorters for perfect vortex modes \cite{vaity2015perfect} - a natural mode basis compatible with long-haul OAM transmission in optical fibers \cite{bozinovic2013terabit,ma2023scaling}. The generalized angle-OAM Talbot effect and its application in mode sorting set as landmark demonstrations which strongly motivate the exploration of the duality between time-frequency and angle-OAM (Supplementary Note 1), for unveiling new phenomena and applications in the realm of OAM and vice versa.

\section*{Results} 

 \begin{figure*}[htp]
  \centering{
  \includegraphics[width = 0.9\textwidth]{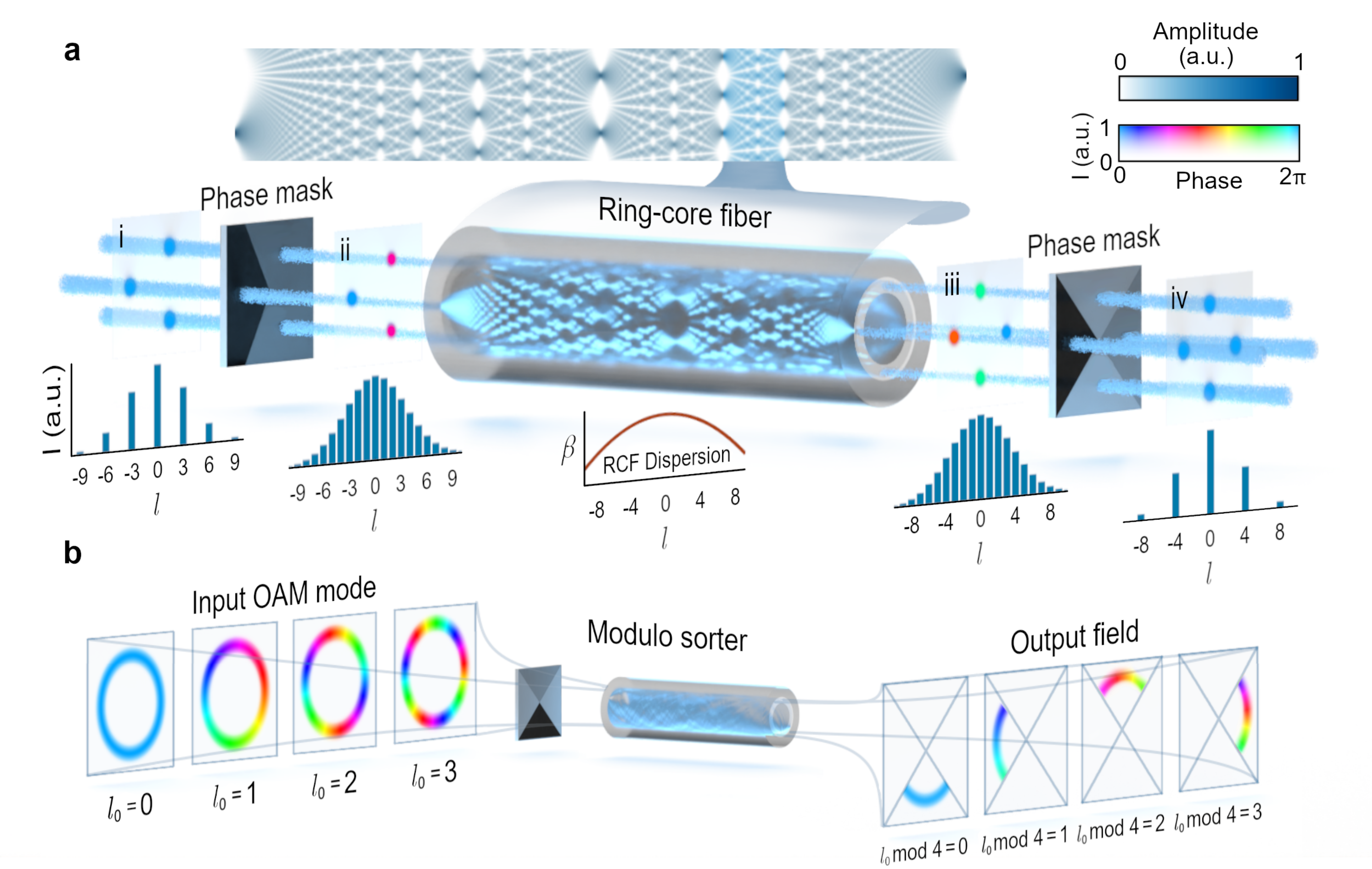}
  } \caption{\noindent\textbf{Schematic diagram of the generalized angle-OAM Talbot effect and modulo OAM mode sorting.} \textbf{a}, Generalized Talbot effect for arbitrary transformations of angular petal numbers and OAM mode spacings of light fields (from $n$ to $m$, $n=3$ and $m=4$ are shown). An optical field consisting of $n$ in-phase petals and an OAM spacing of $n$ impinges on a phase mask, which renders Talbot phases on the angular petals (i to ii). 
    The field is then coupled to a ring-core fiber (RCF) which, upon propagation, induces a quadratic phase relation on the OAM modes (RCF dispersion in the inset), and transforms the field into $m$ angular petals following the pattern evolution in a Talbot carpet (ii to iii). 
    After passing through a second phase mask, the Talbot phases on the $m$ petals are removed and results in an OAM spacing of $m$ (iii to iv). The complex angular fields and their corresponding OAM spectra at each stage are illustrated. 
    \textbf{b}, Talbot-based mode sorter which sorts OAM modes dependent on their orders $l_0$ modulo a predefined integer $q$ ($q =4$, $l_0 =0,1,2,3$ are shown). The sorter consists in a Talbot phase mask and a piece of RCF, and is designed according to the principle of the generalized angle-OAM Talbot effect. Input OAM orders incongruent modulo $q$ are mapped to different, non-overlapping angular sectors at the output of the mode sorter.}
  \label{fig1}
\end{figure*} 

\noindent{\textbf{Principle of operation.}} Figure \ref{fig1}a illustrates the schematic diagram of the generalized angle-OAM Talbot effect. We consider an arbitrary integer of angular petals $n$ at the input being coherently transformed to another integer number $m$ at the output ($n, m \in \mathbb{N}_+$, $n=3$ and $m=4$ are shown). These angular petals are equally-spaced and in-phase, leading to $n$ ($m$)-fold periodicity in the azimuthal angle and consequently an OAM mode spacing of $n$ ($m$). The generalized Talbot effect is realized in three consecutive steps, using twice the OAM self-imaging (i to ii, iii to iv) with the angular self-imaging (ii to iii) in between, implemented with Talbot phase masks and RCF propagation, respectively. Such a transformation can be intuitively understood with a Talbot carpet. The first phase mask assigns a Talbot phase relation on the azimuthal petals to match a pattern found in the Talbot carpet wrapped in a cylindrical shape (i to ii). Propagation through the RCF advances the field in the Talbot carpet, yielding a different number of petals at the output plane (ii to iii). The remaining Talbot phase relation of the petals is finally flattened with the second phase mask (iii to iv). 

The angular self-imaging is based on the quadratic phase shaping of the OAM spectrum, as the propagation constants $\beta$ of OAM modes in the RCF depend quadratically on the topological charge $l$ \cite{eriksson2021talbot}, i.e., $\beta(l) \approx k_{\text{co}} - \frac{l^2}{2k_{\text{co}} R^2}$, with $k_{\text{co}}$ the wavevector and $R$ the mean radius of the ring-core. Propagation of a fraction of the Talbot length $\frac{p}{q}z_T$ with coprime $p$ and $q$ ($p, q \in \mathbb{N}_+$) leads to $q$-fold self-imaging in the azimuthal angle, where $z_T = 2\pi k_{\text{co}} R^2$ is the Talbot length of the RCF at which the inverted integer self-image is reproduced \cite{azana2001temporal}. For the generalized angle-OAM Talbot effect, the RCF length used to transform $n$ petals into $m$ petals can be flexibly chosen by the distance between positions of arbitrary $n$ self-images and $m$ self-images in the Talbot carpet. While the number of self-images is tailored by the angular Talbot effect, conversions between patterns in the Talbot carpet and in-phase petals at the input and output are realized by self-imaging processes in the OAM domain.
The OAM Talbot effect can be seen as the change of the mode spacing in the OAM spectrum, mediated by the phase modulation of angular petals using phase masks with discretely stepped angular Talbot phases \cite{lin2021spectral}. As such, Talbot phase masks with $n$ and $m$ angular sectors are used before and after the RCF for an arbitrary $n$ to $m$ beam splitter/combiner in the polar coordinates.

The generalized angle-OAM Talbot effect also opens up a new OAM sorting scheme based on the Talbot phase modulation and RCF propagation, as depicted in Fig. \ref{fig1}b. Perfect vortex modes with different OAM indices $l_0$ ($l_0 \in \mathbb{Z}$) at the input are projected to $q$ ($q \in \mathbb{Z}_+$ and can be designed arbitrarily) non-overlapping angular sectors at the output of the sorter ($q= 4$ and $l_0 = 0, 1, 2, 3$ are shown). For each input OAM mode, the mapping to the output angular sector is contingent on the remainder of the topological charge $l_0$ modulo $q$. The convergence of light intensity from uniform distributions in the azimuthal angle into angular sectors is analogous to the concept of Talbot array illuminators in space \cite{lohmann1990making} or time \cite{fernandez2017cw}, which transforms a plane wave into an array of illuminators or shapes a continuous-wave (CW) laser into a pulse train, respectively. 

From the generalized self-imaging point of view, the proposed sorting process can be understood as combining $q$ angular petals into one, where, in this case, the petals are rectangular-shaped with an angular duty cycle of $\frac{1}{q}$ and no Talbot phase mask is needed at the RCF output. For a $q$ to $1$ combiner, the RCF should take the length of $\frac{p}{q}z_T$, and the input Talbot phase mask can be designed accordingly with $q$ sectors (indexed as $k = 0,1,...,q-1$) corresponding to quadratic phases of $\theta_k = -\frac{\pi sk^2}{q}$, where $s$ ($s\in \mathbb{N}_+ \cap [1,2q-1]$) is uniquely determined by $p$ and $q$ \cite{fernandez2017structure,romero2019arbitrary,cortes2016generality} ($s=1$ and $p= 1$ are used in  Fig. \ref{fig1}b). This creates a sinc-shaped OAM comb centered at the input mode index $l_0$ and a zero-crossing bandwidth of $2q$, while the phases among these OAM orders follow a Talbot phase relation \cite{lin2021spectral,fernandez2017cw} (Supplementary Note 2). With subsequent RCF propagation, the quadratic phases within the OAM comb can be perfectly cancelled, leading to the construction of a rectangular-shaped sector at the output with an angular width of $\frac{2\pi}{q}$. 
For different input OAM modes, similar angular focusing behaviours are obtained but experience additional rotations depending on the input OAM orders (Fig. \ref{fig1}b). 
The relative rotation is linked to the remaining linear phase relation of the output OAM spectrum in the sorting process, where the output field is rotated counterclockwise by an angle of $2\pi \frac{p}{q} \Delta l$ for an input OAM shift of $\Delta l$ ($\Delta l \in \mathbb{Z}$) (Supplementary Note 2). Since $p$ and $q$ are coprime, the rotation of the output angular sector is periodic in OAM with a period of $q$, whereas OAM orders incongruent modulo $q$ are imaged onto different sectors. Therefore, the system acts as a modulo $q$ OAM sorter. More details on the design rules, the theoretical foundation as well as simulated carpets of Talbot-based sorters are detailed in Supplementary Note 2 and 3.

 \begin{figure*}[htp]
  \centering{
  \includegraphics[width = 0.9\textwidth]{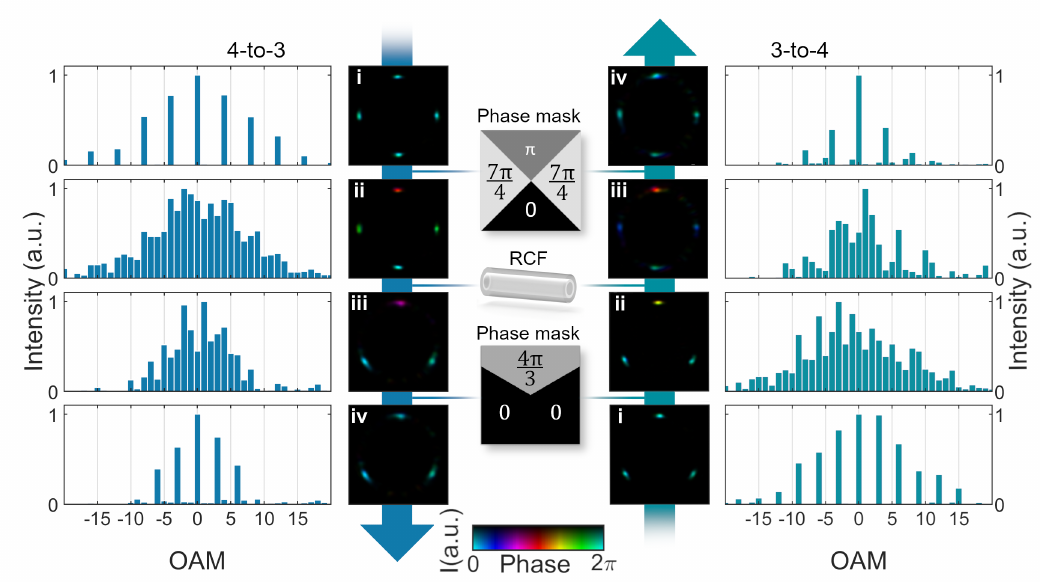}
  } \caption{\noindent\textbf{Experimental demonstration of the generalized angle-OAM Talbot effect.} Conversions of angular petals and OAM spacings from 4 to 3 (left) and 3 to 4 (right). Measured optical fields at (i) Input: 4(3) in-phase angular petals and an OAM spacing of 4(3); (ii) 4(3) angular petals with Talbot phases and an OAM spacing of 1; (iii) 3(4) angular petals with Talbot phases and an OAM spacing of 1; (iv) Output: 3(4) in-phase angular petals and an OAM spacing of 3(4). Phase masks with angular Talbot phase distributions and an RCF in between with a length of $\frac{7}{12}z_T$ (middle) where $z_T$ is the Talbot length, are used to realize these transformations.} 
  \label{fig2}
\end{figure*} 

\noindent\textbf{Generalized angle-OAM Talbot effect.} The experimental setup (Supplementary Note 4) mainly consists of a tunable CW laser in the telecom C-band, two phase-only spatial light modulators (SLMs), RCF pieces and a camera. 
The laser beam illuminates the first SLM, where amplitude and phase modulated holograms are used to generate the petal structures and implement the first Talbot phase mask \cite{Bolduc:2013aa}. Subsequently, the optical field is coupled to a piece of RCF and, after out-coupling, is then directed to the second Talbot phase mask implemented with the other SLM to derive the output field. By recording the interferograms of the optical fields and a reference beam in a camera, off-axis digital holography is used to reconstruct the complex amplitudes of the fields and OAM spectra at different parts of the setup \cite{GoodmanJ.W.1967Diff, verrier2011off}. The RCF employed in the experiment consists of an inner silica (SiO$_2$) cladding with a diameter of 54.7 $\mu$m, a GeO\textsubscript{2}-doped, 2.15 $\mu$m thick ring-core region, and a SiO$_2$ outer cladding \cite{eriksson2021talbot} (Supplementary Note 5). The Talbot length of this RCF is estimated as $z_T \approx 3.00$ cm at the optical wavelength around 1550 nm. 

As a first demonstration, we prepare an RCF piece with a length of $\frac{7}{12}z_T \approx 1.75$ cm, while matching the exact fraction of the Talbot length is achieved by slightly tuning the laser wavelength (Supplementary Note 5). The Talbot phase masks are designed accordingly, rendering the angular Talbot phase at the input and cancelling it at the output (Supplementary Note 6). The complex optical fields are measured experimentally at four planes in Fig. \ref{fig2} (Supplementary Note 4): input (i), after the first phase mask (ii), after the RCF (iii), and after the last phase mask (iv). It can be seen that the 4 in-phase petals are transformed into 3 in-phase petals, and vice versa by swapping the input and output Talbot phase masks.
the input and output Talbot phase masks. 
We also retrieve the OAM spectra of petal fields by calculating the overlap integrals of the measured complex fields and phase vortices, where the mode spacings of the measured OAM spectra confirm again the transformations.

 \begin{figure*}[htp]
  \centering{
  \includegraphics[width = 0.9\textwidth]{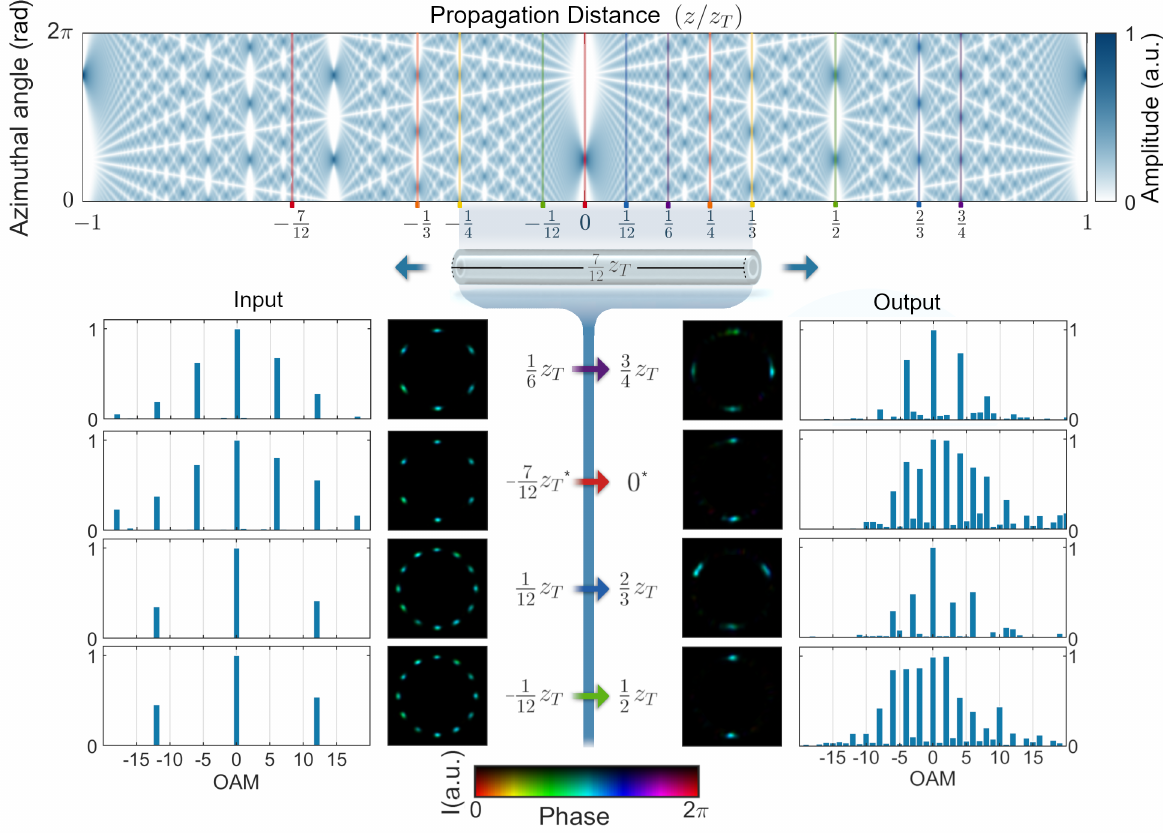}
  } \caption{\noindent\textbf{Versatile angular petal transformations based on the generalized Talbot effect.} The transformations of the petal structures and OAM spectra can be designed with the help of a conventional Talbot carpet (top). Input and output planes are chosen from the Talbot carpet according to the desired numbers of input and output petals, which determines the RCF length and the Talbot phase masks required for the conversion. For a fixed RCF length of $\frac{7}{12}z_T$ shown here, the fiber can slide along the Talbot carpet to realize different transformations. Versatile conversions from 6 to 4 petals, 6 to 2 petals, 12 to 3 petals, and 12 to 2 petals (from top to bottom) are experimentally demonstrated.  Note that the conversion from 6 to 2 petals is the superposition of two 12 to 1 transformations with a $\pi$ shift in the azimuthal angle, which leads to 2 angular petals at the output and only 6 input petals due to interference. The vertical colored lines in the carpet denote the input and output planes of different transformations. The orange- and yellow-colored transformations are seen in Fig. \ref{fig2}, and the rest are shown here. Left: Measured input fields and OAM spectra. Right: Measured output fields and OAM spectra.} 
  \label{fig3}
\end{figure*} 

The ideal angular self-imaging process assumes the orthogonality of the underlying OAM modes. However, modal coupling effects do exist in the RCF if the effective refractive indices of eigenmodes are close, provided by perturbations in the fiber (e.g., bends, ellipticity) and more crucially, the non-ideal mode launching conditions which mediate energy transfer among modes \cite{ma2020propagation, ma2023scaling}. Modal coupling effects are especially prominent for lower-order OAM modes with close effective refractive indices, as the effective refractive index is quadratically dependent on the OAM mode index in the RCF (Fig. \ref{fig1}a).
Furthermore, coupling between phase-matched OAM modes of opposing orders ($l$ and $-l$) modes is frustrated at higher OAM due to OAM conservation.
For these reasons, the first Talbot phase mask is additionally superposed with an OAM vortex of 15, shifting the entire OAM spectrum of the input field to higher OAM modes which are less vulnerable to mode coupling. This OAM shift is undone with the second phase mask, simply with the superposition of the complementary OAM vortex of -15. For clarity, such OAM displacements are omitted when measuring the complex fields at stages (i) and (ii) in Fig. \ref{fig2}. Along with coupling between OAM modes, the OAM spectral content of the field might be perturbed by polarization coupling (between left-handed polarization and right-handed polarization) in the RCF together with the polarization-selectivity of the SLM, and OAM-dependent mode launching and collection efficiencies. These effects are a composite outcome of experimental imperfections as well as modal coupling effects in propagation through the RCF, resulting in the narrowing of the OAM spectral envelopes between stages (ii) and (iii) in Fig. \ref{fig2}.

The generalized angle-OAM Talbot effect can be utilized as a means to perform transformations between fields with arbitrary numbers of petals. Realistically, the transformations are only limited by the OAM spectral width supported by the RCF with minimal modal coupling. While such transformations generally require the use of different RCF lengths, we show that a single RCF piece can also enable versatile angular petal conversions.
We employ the same RCF piece of $1.75$~cm to perform a few more conversions, which can be seen as an RCF sliding along the Talbot carpet (Fig. \ref{fig3}), connecting different input and output planes. This RCF length bridges planes in the Talbot carpet between $\frac{1}{6}z_T$ and $\frac{3}{4}z_T$, $\frac{1}{12}z_T$ and $\frac{2}{3}z_T$, $-\frac{1}{12}z_T$ and $\frac{1}{2}z_T$, enabling conversions from 6 to 4 petals, 12 to 3 petals, and 12 to 2 petals, respectively. Figure \ref{fig3} shows the experimentally characterized petal structures and corresponding OAM spectra at both the input and output of these transformations. Further, we note that multiple azimuthally shifted transformations can be superposed. As an example, the conversion from 6 to 2 petals is a superposition of two 12 to 1 transformations with a $\pi$ shift in the azimuthal angle. Such a superposition renders 2 angular petals at the output, while the interference of the input fields reduces the number of input petals from 12 to 6. The Talbot phase masks used for all the transformations in Fig. \ref{fig3} are illustrated in Supplementary Note 6. 

 \begin{figure*}[htp]
  \centering{
  \includegraphics[width = 0.9\textwidth]{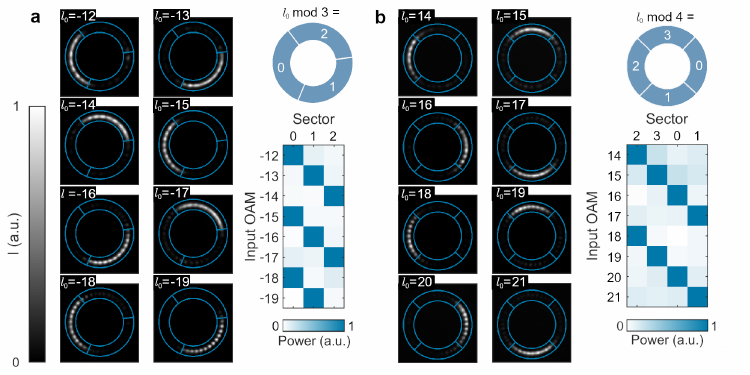}
  } \caption{\noindent\textbf{Talbot-based modulo 3 and modulo 4 OAM mode sorting.} \textbf{a}, Experimental results of the modulo 3 OAM sorter, constructed based on a Talbot phase mask with 3 sectors and an RCF piece with a length of $\frac{2}{3}z_T$. 
    \textbf{b}, Experimental results of the modulo 4 OAM sorter, constructed based on a Talbot phase mask with 4 sectors and an RCF piece with a length of $\frac{3}{4}z_T$. \textbf{a},\textbf{b}, 
    Left: Measured intensity distributions at the sorter output for input OAM modes $l_0$ labeled on the top left corners. Top right: Layout of the modulo sorter. Neighboring OAM modes are mapped to adjacent angular sectors at the sorter output. Bottom right: Crosstalk matrix of the modulo sorter, with on average $86.5\% \pm 3.8\%$ and $72.0\% \pm 5.9 \%$ of optical power being mapped onto the expected angular sectors for the modulo 3 and modulo 4 sorters, respectively. } 
  \label{fig4}
\end{figure*} 

\noindent \textbf{Talbot-based OAM sorter.} As previously stated, the generalized angle-OAM Talbot effect can be harnessed for the application of OAM mode sorting. To illustrate the sorting phenomenon, we prepare RCF pieces with lengths of $\frac{2}{3}z_T \approx 2.00$~cm and $\frac{3}{4}z_T \approx 2.25$~cm, and together with appropriate Talbot phase masks (Supplementary Note 6) to implement modulo 3 and modulo 4 OAM sorters. Again, we use higher-order OAM modes to avoid modal crosstalk between different OAM modes. In practice, sorting of lower-order OAM modes is feasible simply by shifting the input OAM mode with an additional phase vortex incorporated into the Talbot phase mask. Figure \ref{fig4} shows the measured output fields and crosstalk matrices of the modulo 3 and modulo 4 OAM sorters, with input OAM orders ranging from $-19$ to $-12$ and from $14$ to $21$, respectively. The Talbot-based OAM sorter works for both negative and positive OAM orders, achieving excellent sorting performance as manifested in the measured crosstalk matrices, where the relative power distribution is calculated by integrating the measured pixel values of the camera in each sector over the entire annular ring. We calibrate the camera response for accurate evaluation of the sorter performance (Supplementary Note 7). For incident OAM modes shown in Fig. \ref{fig4}, on average $86.5\% \pm 3.8\%$ and $72.0\% \pm 5.9 \%$ of optical power (the error denotes the standard deviation across input OAM orders) is imaged onto the expected angular sectors for the modulo 3 and modulo 4 sorters, respectively. With the increment of the input OAM order, the output field of the modulo 3 (4) sorter shown in Fig. \ref{fig4} is rotated $\frac{4\pi}{3}$ ($\frac{3\pi}{2}$) counter-clockwise in the azimuthal angle, mapping neighboring input OAM orders to adjacent angular sectors and matching with the theoretical prediction based on the RCF length used (Supplementary Note 2).

 \begin{figure*}[htp]
  \centering{
  \includegraphics[width = 0.9\textwidth]{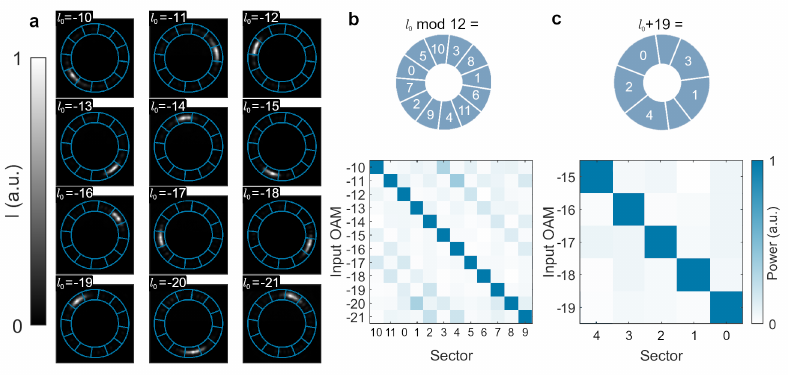}
  } \caption{\noindent\textbf{Unconventional modulo 12 OAM mode sorting and enhanced sorting of 5 consecutive OAM modes.} The mode sorters are constructed based on a Talbot phase mask with 12 sectors and an RCF piece with a length of $\frac{7}{12}z_T$. \textbf{a}, Measured intensity distributions at the sorter output for input OAM modes $l_0$ ranging from -21 to -10 (labeled on the top left corners). \textbf{b}, Top: Layout of the sorter for 12 OAM modes. Neighboring OAM modes are mapped to 7 angular sectors apart. Bottom: Crosstalk matrix of the mode sorter, with an average sorting accuracy of $59.5\pm 6.4\%$ degraded by light leakage to neighboring sectors. \textbf{c}, Since any 5 consecutive OAM modes are never imaged to neighboring sectors, a special set of wider sectors can be constructed to sort these modes with a higher accuracy. Top: Layout of the sorter for OAM orders from -19 to -15. Bottom: Crosstalk matrix of the mode sorter, with an average sorting accuracy increased to $82.8\% \pm 2.6\%$.} 
  \label{fig5}
\end{figure*} 

Figures \ref{fig5}a-b present the experimental results of a Talbot-based OAM sorter with an increased dimensionality, which sorts 12 OAM modes simultaneously in an unconventional way. To achieve this, we employ the same RCF piece used for the generalized angle-OAM Talbot effect, together with a 12-sector Talbot phase mask matched to the chosen fiber length of $\frac{7}{12}z_T$ (Supplementary Note 6). For input OAM orders ranging from -21 to -10, the measured output fields of the mode sorter are depicted in Fig. \ref{fig5}a and their corresponding crosstalk matrix is shown in Fig. \ref{fig5}b. In this scenario, an average of $59.5\pm 6.4\%$ of the optical power is mapped to the expected sector at the output of the sorter. Compared to the sorters above, the experimental performance here is decreased with light leaking to neighboring sectors, due to the increased number of modes to be sorted altogether. Strikingly, the designed sorter projects every consecutive OAM mode not to the neighboring sector as before, but rather 7 sectors away, nearly to the complete opposite side of the ring-core. This is attributed to the used RCF length of $\frac{7}{12}z_T$, causing the output field to rotate counter-clockwise by $\frac{7}{12}\times 2\pi = \frac{7\pi}{6}$ with the increment of the OAM order (Supplementary Note 2). As such, any 5 consecutive OAM modes are never imaged to neighboring sectors, and this peculiar property can be leveraged to enhance the sorter performance at the cost of a reduced measurement space. For example, a special set of 5 wider sectors, twice as wide as the original ones with buffer zones at both sides of the ideal output positions, can be arranged to sort 5 neighboring OAM orders from -19 to -15 (Fig. \ref{fig5}c), thereby increasing the sorting accuracy to $82.8\% \pm 2.6\%$. 
The accuracy can be further improved by narrowing the width of the wide sectors. Considering the central $90\%$ ($75\%$) of all wide sectors yields $84.3\% \pm 2.3\%$ ($86.5\% \pm 1.8\%$) of measured power in the expected sector.
While reducing the crosstalk, it must be emphasized that considering a portion of the wide sector is an inherently lossy procedure, since part of the light is not considered in the assessment. As another example, a sorter for 9 modes based on an RCF length of $\frac{7}{9}z_T$ also maps neighboring OAM orders to non-adjacent angular sectors (Supplementary Note 8).

\vspace{0.1cm}

\section*{Discussion} 
\noindent{While} space-time duality is widely known to link physical phenomena in the spatial and temporal domains, the temporal analogues in space are predominantly manifested in the Cartesian coordinates \cite{kolner1994space}. Here, we extend such connections to the azimuthal angle in the polar coordinates, and identify the duality among angle-OAM, time-frequency and position-momentum (Supplementary Note 1). Consequently, diverse temporal and spectral phenomena are also present in the angular and OAM forms, leading to the observation of the generalized angle-OAM Talbot effect in this study. The RCF used in the experiment supports a wide range of OAM orders as well as a number of low-loss leaky OAM modes \cite{ma2023scaling} (Supplementary Note 5). This allows for the realization of very high-order beam splitters and combiners, also potentially achieving passive amplification of angular petals and OAM modes \cite{romero2019arbitrary,maram2014noiseless}.

The OAM sorting scheme demonstrated in this work, unlike all the existing sorters, utilizes the generalized Talbot effect and is in principle free from any crosstalk within a predefined measurement range. So far, most of the scalable OAM sorters are fundamentally subjected to the modal crosstalk. This includes methods based on the log-polar transformations \cite{berkhout2010efficient,lavery2012refractive}, as well as their improved variants using beam copying \cite{mirhosseini2013efficient} and spiral transformation \cite{wen2018spiral}, which still face challenges to completely separate neighboring OAM modes. In comparison, the Talbot-based sorters can be designed to steer neighboring OAM modes to almost furthest positions in a circle (Fig. \ref{fig5}). The modulo sorting behaviour of Talbot-based OAM sorters is similar to sorters built upon cascaded Mach–Zehnder interferometers with Dove prisms \cite{leach2002measuring} but in a more scalable manner. 
While sorters based on multi-plane light conversion are task-agnostic \cite{fontaine2019laguerre}, the multi-plane phase patterns are obtained with iterative computer optimizations and thereby less interpretable. Conversely, our OAM sorters can be designed on demand based on the generalized self-imaging effect, and achieve excellent experimental performance on par with the state of the art sorters \cite{mirhosseini2013efficient,wen2018spiral}. Although being crosstalk-free theoretically, the practical performance of the Talbot-based sorters is constrained by several factors in the experiment. The primary issue arises from various modal coupling mechanisms and mode launching imperfections, which transfer energy among different OAM orders and lead to light illuminating incorrect sectors at the output. Moreover, since the field after the Talbot phase mask corresponds to a sinc-shaped OAM spectrum, the higher-order OAM components not supported by the RCF are filtered, resulting in the OAM spectral narrowing and washing out the steep cutoff between different angular sectors. While solving the technical limitations can be challenging, propagation of OAM modes with negligible crosstalk can be achieved with careful mode launching and fiber design \cite{ma2023scaling, ma2020propagation}. Moreover, the method presented in Fig. \ref{fig5} could enhance the sorting performance even with the experimental imperfections in place. 

In summary, we experimentally demonstrate the generalized Talbot effect in the angle-OAM domain and utilize the principle for efficient OAM mode sorting. The observed effect and applications are not only limited to optical waves, but may be extended to other types of waves including acoustics and matter waves \cite{nowak1997high}. The Talbot-based mode sorters are essentially gratings for OAM modes, serving as a building block for OAM division multiplexing systems. Similar to a Fourier-domain pulse shaper \cite{weiner2000femtosecond}, the Talbot-based sorter empowers light shaping in the azimuthal angle via mode-by-mode shaping of each individual OAM component. Besides classical applications, the arbitrary higher-order beam splitting demonstrated in this work could be of interest in quantum optics, e.g., for linear optical networks employed in quantum information processing \cite{flamini2018photonic, carolan2015universal} and multiphoton interferometry \cite{PanJian-Wei2012Meai, KumarShreya2023Eegu}. The observed generalized angle-OAM Talbot effect is just one of the examples of the duality between angle-OAM and time-frequency, suggesting that the phenomena and processing techniques in these two fields are broadly connected.

\vspace{0.5cm}

\medskip
\begin{footnotesize}

\noindent \textbf{Acknowledgements}:
This work was supported by European Research Council (TWISTION, 101042368). J.H. acknowledges Swiss national science foundation fellowship (P2ELP2$\_$199825). S.G. is a member of the institut Universitaire de France. M.E. acknowledges the Research Council of Finland Flagship Programme, Photonics Research and Innovation (PREIN), 320165. R.F. acknowledges the Research Council of Finland through the Academy Research Fellowship (Decision 332399).

\vspace{0.1cm}

\noindent \textbf{Data Availability Statement}: 
The data and code that support the plots within this paper and other findings of this study are available from the corresponding author upon reasonable request.
\end{footnotesize}

\renewcommand{\bibpreamble}{
$^\dagger$These authors contributed equally to this work.\\
$^\ast${Corresponding author: \textcolor{magenta}{jianqi.hu@lkb.ens.fr}}\\
}

\bibliographystyle{naturemag}
\bibliography{ref}

\end{document}


\title{Supplementary information for:\\ Generalized angle-orbital-angular-momentum Talbot effect and modulo mode sorting}

\author{Jianqi Hu$^{1,\dagger,*}$, Matias Eriksson$^{2,\dagger}$, Sylvain Gigan$^{1}$ \& Robert Fickler$^{2}$}
\affiliation{
$^1$Laboratoire Kastler Brossel\char`,{} École Normale Supérieure - Paris Sciences et Lettres (PSL) Research University\char`,{} Sorbonne Université\char`,{} Centre National de la Recherche Scientifique (CNRS)\char`,{} UMR 8552\char`,{} Collège de France\char`,{} 24 rue Lhomond\char`,{} 75005 Paris\char`,{} France.\\
$^2$Physics Unit\char`,{} Photonics Laboratory\char`,{} Tampere University\char`,{} 33720 Tampere\char`,{} Finland.\\
}
\maketitle

\section*{\textbf{Supplementary Note 1. Duality between time-frequency and angle-OAM}}

In Supplementary Fig. \ref{fig_s1}, we outline the duality between the time-frequency and angle-OAM. We elucidate the connections among temporal, spectral, angular, and OAM Talbot effects, as well as the techniques and devices necessary for inducing each of these effects. This is an extension of the previously identified duality between position-momentum and time-frequency for Talbot effects \cite{cortes2016generality}. 

\begin{figure}[htp]
  \renewcommand{\figurename}{Supplementary Figure}
    \centering
    \includegraphics[width=0.6\textwidth]{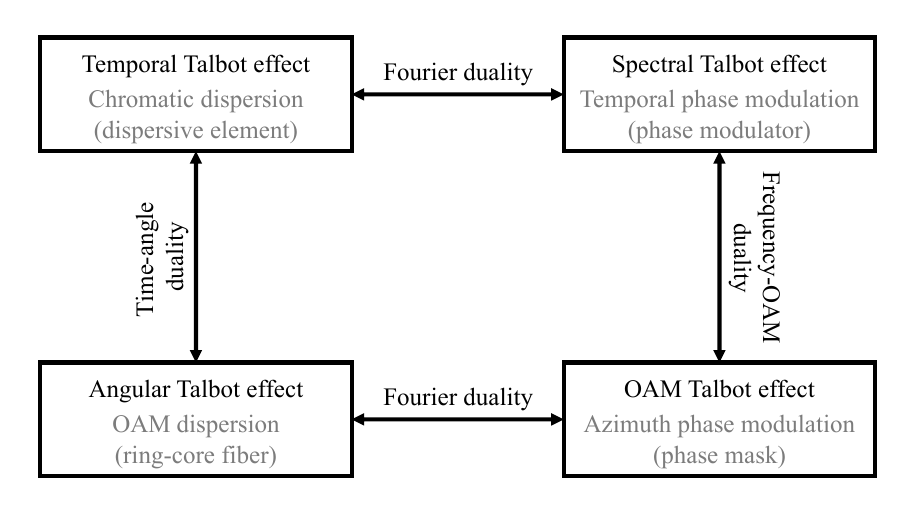}
    \caption{Illustration of the duality between time-frequency and angle-OAM in Talbot effects. Methods and devices necessary for inducing these Talbot effects are also described.}
    \label{fig_s1}
\end{figure}

\section*{\textbf{Supplementary Note 2. Theoretical derivation of Talbot-based OAM sorters}}
In this section, we elucidate the design principle of a Talbot-based OAM sorter, which consists of a Talbot phase mask followed by a section of RCF. To design a modulo $q$ OAM sorter with neighboring OAM orders being spaced by $p$ sectors (center to center) at the output ($p$ and $q$ are coprime), an RCF with $p/q$ of its Talbot length is used. Additionally, the employed Talbot phase mask $T_{q}(\phi) = e^{i\theta(\phi)}$ is composed of $q$ sectors equally divided in the azimuthal angle $\phi$, with each sector ($k = 0,1,...,q-1$) corresponding to a static Talbot phase:
\begin{equation}
\theta_k(\phi)  = - \frac{\pi s k^2}{q} \quad \text{for} \quad \phi \in [\frac{2\pi k}{q},\frac{2\pi(k+1)}{q}),
\label{eq1}
\end{equation}
where $s$ is an integer in the range of $[1,2q-1]$, which can be uniquely determined by $p$ and $q$ \cite{fernandez2017structure}:
\begin{equation}
sp \equiv 
\begin{cases}
\begin{aligned}
1 \pmod {2q}, \quad \mathrm{if} \quad  q\equiv 0 \pmod 2 \\
1+q \pmod {2q}, \quad \mathrm{if} \quad q\equiv 1 \pmod 2\\
\end{aligned}
\end{cases}
\label{eq2}
\end{equation}
A table of $s$ value as a function of different pairs of $p$ and $q$ can be found in Refs. \cite{cortes2016generality,fernandez2017structure}. $s$ and $q$ are also coprime and have the opposite parity \cite{romero2019arbitrary}. In the following, we prove that the choices of the Talbot phase mask and RCF length given above act as a modulo q OAM sorter with $p$-sector separation for neighboring OAM orders. Consider an OAM mode $e^{il_0 \phi}$ is input to the designed system, the output OAM spectrum after the Talbot phase mask is given by:
\begin{equation}
\begin{aligned}
A_{\text{mask}}[l] & =  \frac{1}{2\pi}\int_{0}^{2\pi} T_q(\phi) e^{il_0\phi} e^{-il\phi} \,d\phi \\
& = \frac{1}{2\pi} \sum_{k=0}^{q-1} \int_{2\pi k/q}^{2\pi (k+1)/q} e^{i \theta_k -i(l-l_0)\phi} \,d\phi \\
& = \frac{1}{2\pi} \sum_{k=0}^{q-1} e^{i\theta_k} \frac{e^{-i(l-l_0)\phi}}{-i(l-l_0)}\bigg|_{2\pi k/q}^{2\pi (k+1)/q} \\
& = \frac{i}{2\pi (l-l_0)} \sum_{k=0}^{q-1} e^{i \theta_k} e^{-i2\pi(l-l_0) k/q} ( e^{-i2\pi(l-l_0)/q} -1) \\
& = \frac{\mathrm{sinc}(\frac{l-l_0}{q})}{q} e^{-i\pi(l-l_0)/q} \sum_{k=0}^{q-1} e^{-i \pi s k^2/q} e^{-i2\pi(l-l_0) k/q},
\end{aligned}
\label{eq3}
\end{equation}
where $\mathrm{sinc}(x) = \frac{\sin(\pi x)}{\pi x}$. $A_{\text{mask}}[l]$ denotes the complex amplitude of OAM mode $l$ at the output of the mask. To calculate the summation $\sum_{k=0}^{q-1} e^{-i \pi s k^2/q} e^{-i2\pi(l-l_0)  k/q}$ in the last equality of Eq. \eqref{eq3}, we examine separately for $q$ being even and odd. 

\noindent{1) $q \equiv 0 \pmod 2$.} \\ 
\indent{For} an even $q$, we note that changing the running variable $k$ to $pk'$ with $k' = 0,...,q-1$ will not change the sum \cite{fernandez2017structure}. This is justified by the following two reasons: 1) $pk'$ also forms a complete residue system of $q$, the same as $k = 0,..., q-1$ because $p$ and $q$ are coprime; 2) if $pk' \equiv k \pmod q$, the summation term $e^{-i \pi s (pk')^2/q} e^{-i2\pi(l-l_0) pk'/q}$ is equal to $ e^{-i \pi s k^2/q} e^{-i2\pi(l-l_0) k/q}$ for an even $q$. As such, the summation rewrites:
\begin{equation}
\begin{aligned}
\sum_{k=0}^{q-1} e^{-i \pi s k^2/q} e^{-i2\pi(l-l_0) k/q} &= \sum_{k'=0}^{q-1} e^{-i \pi s (pk')^2/q} e^{-i2\pi(l-l_0) pk'/q} \\
&= \sum_{k'=0}^{q-1} e^{-i \pi  pk'^2/q} e^{-i2\pi(l-l_0) pk'/q} \\
&= e^{i \pi p(l-l_0)^2/q} \sum_{k'=0}^{q-1} e^{-i \pi p(k'+l-l_0)^2/q}\\
&= e^{i \pi p(l-l_0)^2/q} \sum_{k'=0}^{q-1} e^{-i \pi pk'^2/q}\\
&= \sqrt{q}\bigg(\frac{q}{p}\bigg) e^{-i\pi p/4}e^{i \pi p(l-l_0)^2/q},
\end{aligned}
\label{eq4}
\end{equation}
where $\big(\frac{a}{b}\big)$ is the Jacobi symbol, which is $+1$ if there exists an integer $n$ such that $n^2 \equiv a \pmod b$, otherwise $-1$. In Eq. \eqref{eq4}, the second equality is derived using $sp \equiv 1 \pmod {2q}$, where the last equality is obtained using an identity of generalized quadratic Gauss sums \cite{fernandez2017structure}.

\noindent{2) $q \equiv 1 \pmod 2$.} \\ 
\indent{For} an odd $q$, the summation is calculated in a similar way as for an even number. In this case, we change the running variable $k$ to $2pk'$ with $k' = 0,...,q-1$. This is valid because: 1) $2pk'$ with $k' = 0,...,q-1$ also runs a complete residue system of $q$, as $2p$ and $q$ are coprime; 2) if $2pk' \equiv k \pmod q$, the summation term $e^{-i \pi s (2pk')^2/q} e^{-i 4\pi(l-l_0) pk'/q}$ is again equal to $ e^{-i \pi s k^2/q} e^{-i2\pi(l-l_0) k/q}$ since $s$ is even for an odd $q$. Thus, the summation in this scenario becomes:
\begin{equation}
\begin{aligned}
\sum_{k=0}^{q-1} e^{-i \pi s k^2/q} e^{-i2\pi(l-l_0)  k/q} &= \sum_{k'=0}^{q-1} e^{-i \pi s (2pk')^2/q} e^{-i4\pi(l-l_0) pk'/q} \\
&= \sum_{k'=0}^{q-1} e^{-i 4\pi (1+q)pk'^2/q} e^{-i4\pi(l-l_0)  pk'/q} \\
&= e^{i \pi p(l-l_0)^2/q} \sum_{k'=0}^{q-1} e^{-i \pi p(2k'+l-l_0)^2/q}\\
&= e^{i \pi p(l-l_0)^2/q} \sum_{k'=0}^{q-1} e^{-i 4\pi pk'^2/q}\\
&= \sqrt{q}\bigg(\frac{4p}{q}\bigg) e^{i\pi (q-1)/4}e^{i \pi p(l-l_0)^2/q},
\end{aligned}
\label{eq5}
\end{equation}
where we use $sp \equiv 1+q \pmod{2q}$ and an identity from Ref. \cite{fernandez2017structure} to derive Eq. \eqref{eq5}.

After analyzing separately with respect to $q$'s parity, the complex amplitude of OAM mode $l$ after the Talbot phase mask can be merged as: 
\begin{equation}
 A_{\text{mask}}[l] = \frac{\mathrm{sinc}(\frac{l-l_0}{q})}{\sqrt{q}} e^{i\varphi_0} e^{-i\pi(l-l_0)/q} e^{i \pi p(l-l_0)^2/q},
\label{eq6}
\end{equation}
where $\varphi_0$ includes constant phases that are global to all the OAM modes once the fiber length is chosen ($p$ and $q$ are fixed). $\varphi_0$ takes the value of $\frac{\pi}{2} (1-\big(\frac{q}{p}\big)) -\frac{\pi p}{4}$ when $q$ is even, and $\frac{\pi}{2} (1-\big(\frac{4p}{q}\big)) +\frac{\pi (q-1)}{4}$ when $q$ is odd. Then the optical field is transmitted through a section of RCF with the length of $\frac{p}{q} z_T$, in which OAM modes pick up propagation phases that are quadratically dependent on the mode indices, i.e., a phase shift of $-\frac{\pi p l^2}{q}$ for the $l$-th OAM mode. As such, the complex amplitude of an OAM mode $l$ at the fiber output is given by:
\begin{equation}
\begin{aligned}
 A_{\text{sorter}}[l] & = \frac{\mathrm{sinc}(\frac{l-l_0}{q})}{\sqrt{q}} e^{i\varphi_0} e^{-i\pi(l-l_0)/q} e^{i \pi p(l-l_0)^2/q} e^{-i\pi p l^2/q}\\
 &=  \frac{\mathrm{sinc}(\frac{l-l_0}{q})}{\sqrt{q}} e^{i\varphi_0}  e^{i\pi(l_0+pl_0^2)/q} e^{-i\pi(1+2pl_0)l/q},
\end{aligned}
\label{eq7}
\end{equation}
where the quadratic phase of the fiber perfectly cancels the OAM-dependent quadratic phase acquired after the mask. 

Several observations can be made from Eq. \eqref{eq7} for a given input OAM mode $l_0$. First,  the OAM spectrum is translation-invariant with respect to the input OAM order $l_0$. It shows a sinc-shaped envelope, which is centered around the OAM mode $l_0$ with the first two zero-crossings at modes $l_0 \pm q$. Second, we look at the phase of each OAM mode at the output for a given input OAM mode $l_0$. As mentioned above, $\varphi_0$ is fixed once $p$ and $q$ are chosen, while the phase $\pi(l_0+pl_0^2)/q$ is global for all OAM modes $l$. The only $l$-dependent phase is $\pi(1+2pl_0)l/q$ which shows a linear dependence. Therefore, the sinc-shaped amplitudes of the OAM spectrum and their linear phase relation correspond to a rectangular amplitude function in the azimuthal angle, i.e., a rectangular-shaped angular sector. The width of the angular sector is $2\pi/q$, equal to a duty cycle of $1/q$ for $2\pi$ angle. This is determined by the OAM spectrum expanded after the Talbot phase mask, whose first two zero-crossings of the sinc envelope are spaced by $2q$. In addition, the angular sector rotates according to the linear phase in the OAM spectrum. It is centered at the azimuthal angle of $-\pi(1+2pl_0)/q$ radian for an input OAM mode $l_0$. 
Thus, the output angular sectors for neighboring OAM modes are rotated by $2\pi p/q$ radians, which are spaced by $p$ sectors in between (center to center). Also, OAM modes congruent modulo $q$ are mapped to the same angular sector, while modes that are incongruent modulo $q$ are directed to different angular sectors since $p$ and $q$ are coprime. 
With all of this, we have a complete explanation of the working principle of the Talbot-based OAM mode sorter.

\begin{figure}[htbp]
  \renewcommand{\figurename}{Supplementary Figure}
    \centering
    \includegraphics[width=0.85\textwidth]{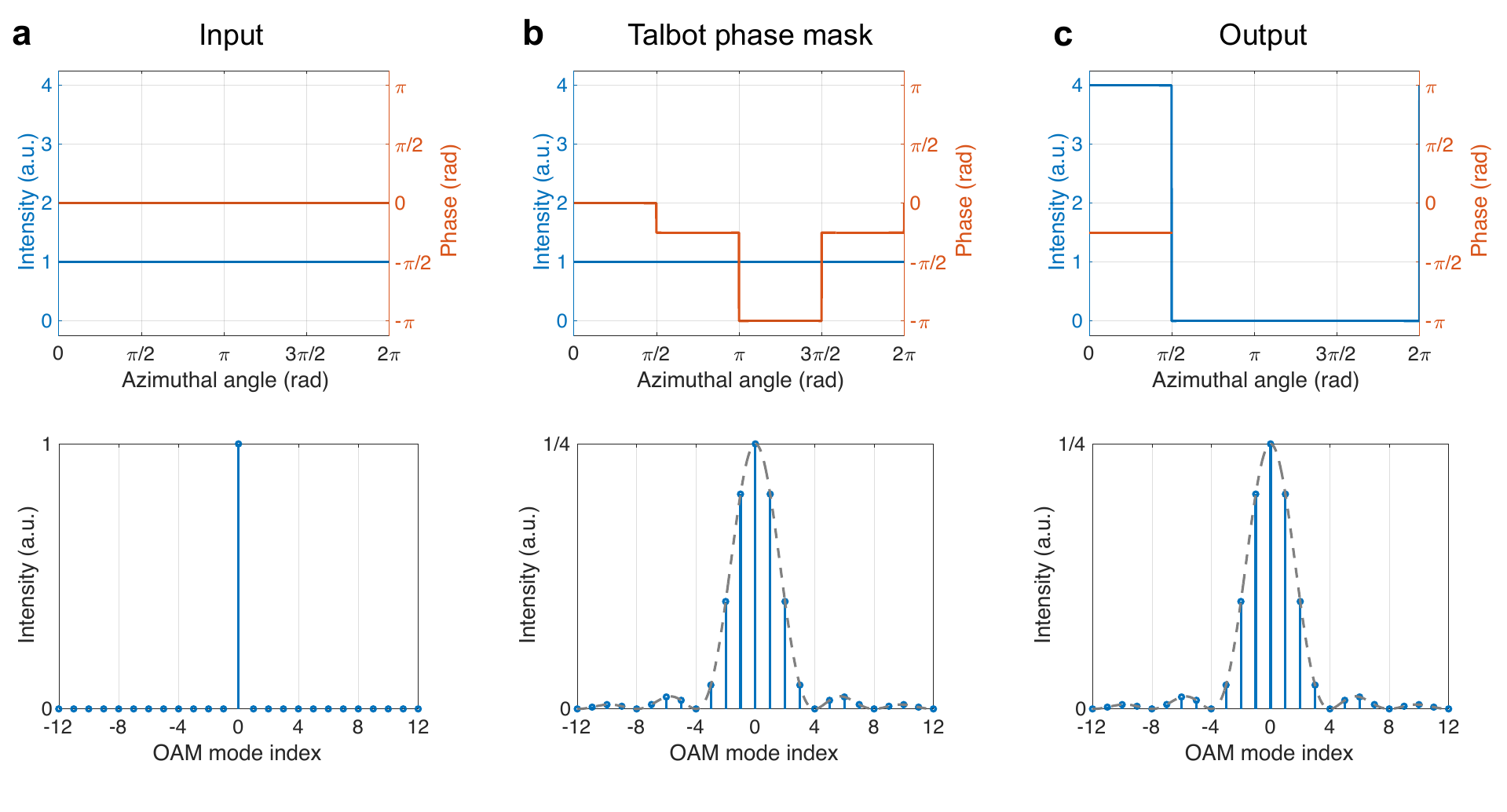}
    \caption{The input, intermediate and output optical fields in a Talbot-based OAM sorter. Parameters $p=1$, $q=4$, and $s =1$ are used, which corresponds to the use of an RCF with a length of $\frac{1}{4}z_T$ and a Talbot phase mask with azimuthal phase distribution of $[0, -\pi/4,-\pi, -\pi/4]$. As an example, an OAM mode $l_0 =0$ is input to the sorter. 
    The intensity and phase distributions of the azimuthal fields (top), and the corresponding OAM spectra are shown \textbf{a} at the input, \textbf{b} after Talbot phase mask, and \textbf{c} at the output. The dashed gray curves in OAM spectra are ideal sinc-squared functions. 
    }
    \label{fig_s2}
\end{figure}

\begin{figure}[!h]
  \renewcommand{\figurename}{Supplementary Figure}
    \centering
    \includegraphics[width=0.85\textwidth]{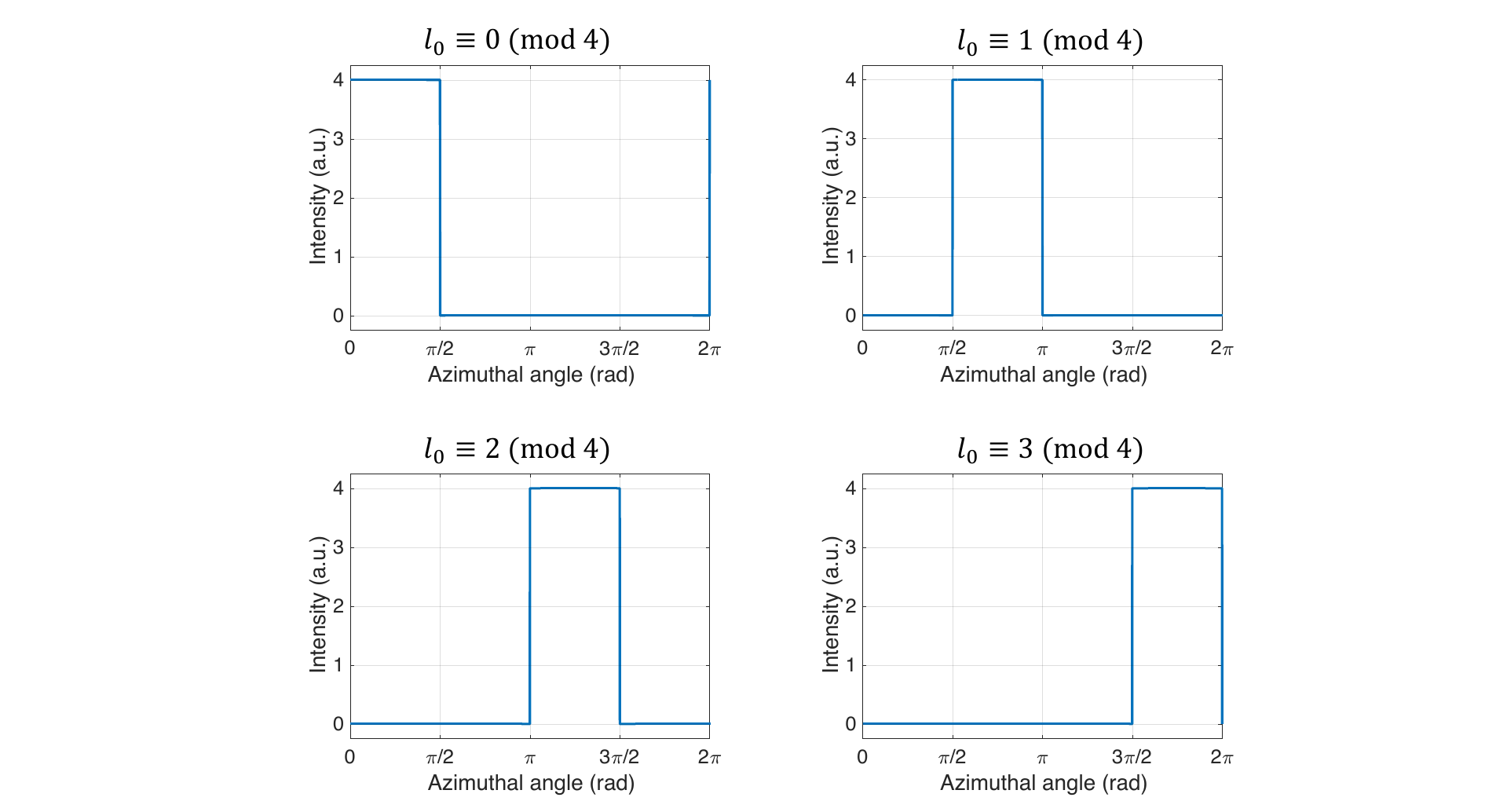}
    \caption{Modulo 4 OAM sorter. OAM modes with remainders of 0, 1, 2, and 3 when modulo 4 are sorted to angular sectors of $[0, \pi/2]$, $[\pi/2, \pi]$, $[\pi,3\pi/2]$, and $[3\pi/2, 2\pi]$, respectively. The design of the Talbot-based OAM sorter is the same as in Supplementary Fig. \ref{fig_s2}.}
    \label{fig_s3}
\end{figure}

In the following, we visualize the proposed OAM sorting scheme with numerical simulations shown in Supplementary Figs. \ref{fig_s2} and \ref{fig_s3}. As an example, we design a modulo 4 OAM sorter using an RCF with a fiber length of $\frac{1}{4}z_T$, which corresponds to one fourth of the Talbot length, i.e., $p=1$ and $q=4$. According to Eq. \eqref{eq2}, we derive $s=1$ such that the four quadrants of the angular Talbot phase mask are modulated with static phases $[0, -\pi/4,-\pi, -\pi/4]$. In Supplementary Fig. \ref{fig_s2}, we simulate the azimuthal fields and OAM spectra at each stage of the sorter. We showcase sending an OAM mode of $l_0 =0$ (Supplementary Fig. \ref{fig_s2}a) to the designed OAM sorter. After being phase modulated by the Talbot phase mask, the optical field remains uniform in intensity but picks up Talbot phases in the azimuthal angle, and its OAM spectral intensity shows a sinc-squared envelope (Supplementary Fig. \ref{fig_s2}b). Then the optical field is further propagated through the RCF. At the output of the fiber, the azimuthal intensity is redistributed from uniform distribution in the $2\pi$ angle to only one quarter of it from $0$ to $\pi/2$ (Supplementary Fig. \ref{fig_s2}c). The output azimuthal field shows an ideal rectangular-shaped distribution, which is resulted from the sinc-shaped amplitude profile and linear phase relation in the OAM spectrum. 

Supplementary Fig. \ref{fig_s3} shows the outputs of the Talbot-based OAM sorter (same as in Supplementary Fig. \ref{fig_s2}) with different OAM mode inputs. When the input OAM modes have the same remainders divided by 4, the sorter maps them to the same angular sectors. This is the reason we refer to the Talbot-based OAM sorter as a modulo sorter. It can be seen that OAM orders with remainders of 0, 1, 2, and 3 when modulo 4 are sorted to angular sectors of $[0, \pi/2]$, $[\pi/2, \pi]$, $[\pi,3\pi/2]$, and $[3\pi/2, 2\pi]$, respectively. As such, for $p=1$, the output field rotates counter-clockwise $\pi/2$ radians in the azimuthal angle with the increment of the input OAM order.

\section*{\textbf{Supplementary Note 3. Talbot carpets for Talbot-based mode sorters}} 
 In Supplementary Fig. \ref{fig_s4}, we show the simulated Talbot carpets for a Talbot-based mode sorter. As an example, we still use the same design of the modulo 4 sorter as in Supplementary Figs. \ref{fig_s2} and \ref{fig_s3}. 
 Supplementary Fig. \ref{fig_s4} illustrates the azimuthal field evolution in an RCF for different input OAM mode indices of $l_0 = 0,1,2,3,4$, after being modulated by the Talbot phase mask at the input of the fiber, i.e., $z=0$. At the propagation distance of $\frac{1}{4}z_T$ in the RCF, the input OAM fields are converged to angular sectors depending on their OAM mode index modulo $4$. Noticeably, although input OAM modes of 0 and 4 show different Talbot carpets, they are indeed sorted to the same angular sector at $z= \frac{1}{4}z_T$. 

\begin{figure}[b!]
  \renewcommand{\figurename}{Supplementary Figure}
    \centering
    \includegraphics[width=0.8\textwidth]{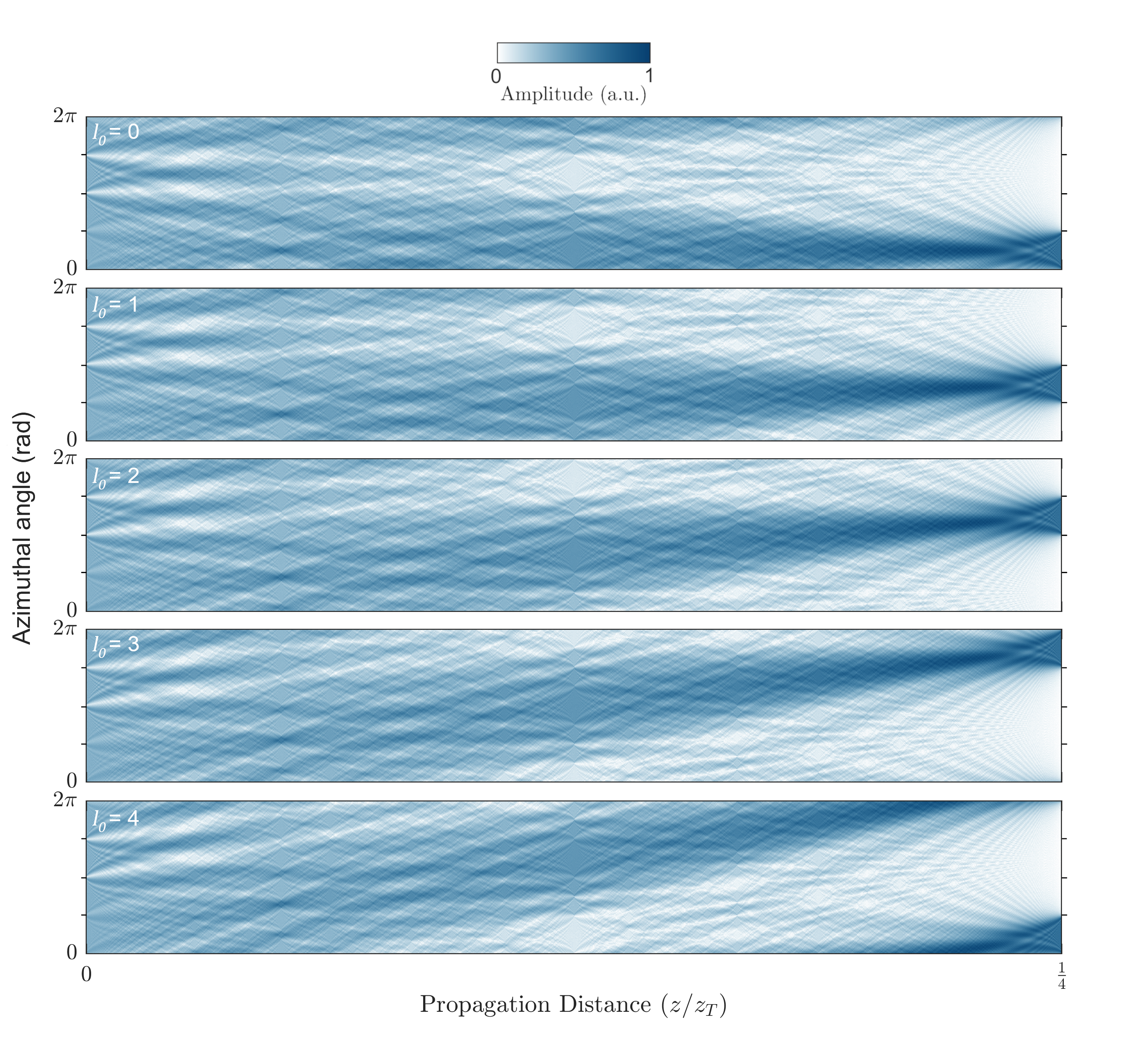}
    \caption{Talbot carpets for a Talbot-based OAM sorter. 
    Simulated azimuthal field evolution in an RCF for input OAM modes of $l_0= 0,1,2,3,4$ (from top to bottom). 
    The input OAM fields are modulated by an angular Talbot phase mask at $z=0$ and propagated through an RCF, in which the mask and fiber length take the same design of the modulo 4 sorter in Supplementary Figs. \ref{fig_s2} and \ref{fig_s3}. At the output plane $z = \frac{1}{4}z_T$, the fields converge to different angular sectors depending on their OAM mode index modulo $4$.} 
    \label{fig_s4}
\end{figure}

\section*{\textbf{Supplementary Note 4. Detailed experimental setup}}
\begin{figure}[h!]
  \renewcommand{\figurename}{Supplementary Figure}
    \centering
    \includegraphics[width=0.95\textwidth]{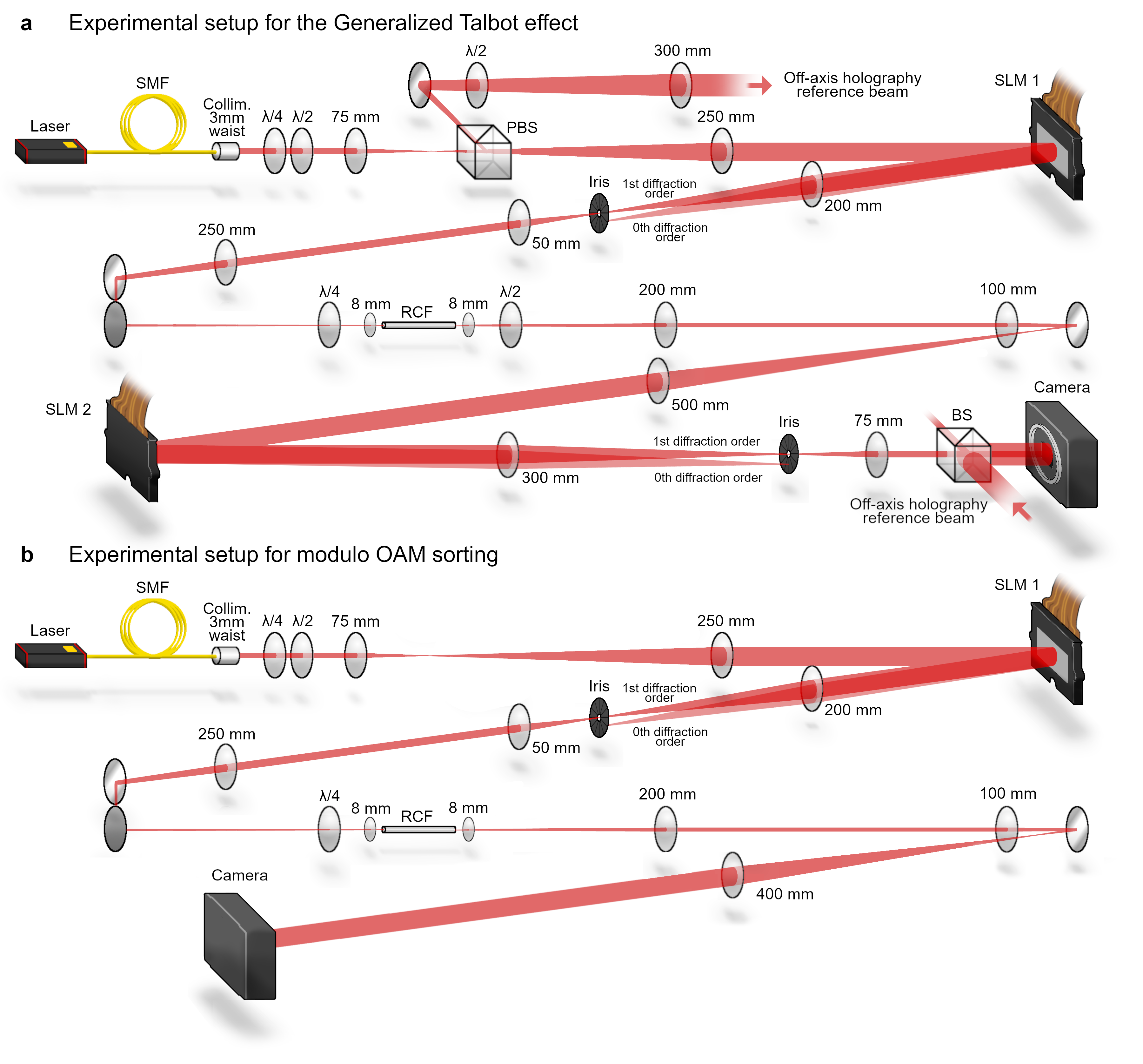}
    \caption{Experimental setup. 
    SMF: single-mode fiber; Collim.: collimator; PBS: polarization beam splitter; BS: beam splitter; $\lambda/4$: quarter-wave plate; $\lambda/2$: half-wave plate.
    \textbf{a} The setup of the generalized angle-OAM Talbot effect experiments, i.e., for Figs. 2 and 3 of the main text. \textbf{b} The setup for modulo OAM sorting experiments, i.e., for Figs. 4 and 5 of the main text.}
    \label{fig_s5}
\end{figure}
Supplementary Fig. \ref{fig_s5}a shows the schematic setup for the generalized angle-OAM Talbot effect experiment. A C-band tunable continuous-wave laser (ID Photonics CoBrite DX1 S) is used as the light source, with wavelength tuning range from 1528 nm to 1568 nm. The laser beam from a single-mode fiber (SMF) through a collimator (beam diameter $3.0~$mm) is magnified via a 4-\textit{f} system (lenses with focal lengths of 75 mm and 250 mm) to illuminate a large part of the first SLM screen (SLM 1, Holoeye Pluto-2). A polarization beam splitter (PBS) along with polarization optics is used to align the beam polarization with the SLM working-axis, and tap part of light as the reference beam for off-axis holography measurement. The light field emerging from SLM 1 then enters a 4-\textit{f} system (200 mm and 50 mm lenses) with 
an iris in its Fourier plane to select the first diffraction order, so as to synthesize optical fields with arbitrary amplitude and phase \cite{Bolduc:2013aa}. At the first SLM, a computer-generated hologram carves out the transverse input angular petals and applies the first Talbot phase mask (see Supplementary Note 6). Through another 4-\textit{f} system (250 mm and 8 mm lenses), this field is imaged onto the input facet of the RCF. A quarter-wave plate (QWP) is placed before the RCF to convert the horizontally polarized light to circularly polarized light, to match the polarization of OAM eigenmodes of the RCF. The light field at the output facet of the RCF is imaged onto the second SLM (SLM 2, Holoeye Pluto-2) with two cascaded 4-\textit{f} systems, where the second SLM applies the second Talbot phase mask (see Supplementary Note 6) superposed with a blazed phase grating. Afterwards, the field is demagnified via a final 4-\textit{f} system, where an iris is placed in the Fourier plane to select the first diffraction order to filter out any unmodulated light. The field is overlapped with an off-axis reference beam in a beam splitter (BS) to record an interferogram of the field on a camera (Point Grey CMLN-13S2M-CS). The off-axis holography measurement retrieves the amplitude and phase information of optical fields at the output \cite{GoodmanJ.W.1967Diff, verrier2011off}. The camera in use possesses a larger number of pixels, which is very useful for resolving the interference fringes of the superposed signal and off-axis reference fields. Although the pixel response of this camera is noisy, the noise consists of very high spatial frequency components and is naturally Fourier-filtered in the off-axis holography field reconstruction process. 

We also use the off-axis holography to measure the fields at planes (i) and (ii), (iii) shown in Fig. 2 of the main text. In reality, the fields at planes (i) and (ii) are measured at the same position, by placing the camera and overlapping the field with the reference beam at the output plane of the first 4-\textit{f} system after SLM 1, with the first Talbot phase mask inactive and active, respectively. Similarly, the fields at planes (iii) and (iv) are measured at the same position with the second Talbot phase mask inactive and active, respectively. 

For accurate phase modulation of the SLMs, the phase masks are superposed with correction terms retrieved by an aberration correction process based on the Grechberg-Saxton phase retrieval algorithm \cite{Jesacher:2007aa}, and further with Zernike polynomials \cite{LakshminarayananVasudevan2011Zpag}. The Zernike coefficients are manually optimized by eliminating asymmetries of the perfect vortex beams generated on the SLMs, and observed in the Fourier plane. The aberration correction terms for SLM 2 are retrieved using a separate, back-propagated laser beam. In the generalized angle-OAM Talbot experiments, we upshift the input OAM spectrum on the SLM 1 with an additional OAM order of $15$ and then downshift on the SLM 2 the complementary order. This is implemented to avoid using OAM modes with $|l|$ close to 0, which have similar effective refractive indices and thus more likely to couple to each other in the RCF \cite{ma2020propagation, ma2023scaling}. 

Supplementary Fig. \ref{fig_s5}b illustrates the setup for the OAM sorting experiment. The main setup is the same as in Supplementary Fig. \ref{fig_s5}a, but without the need for the second SLM and the subsequent imaging setup, and no reference beam is used. Instead, we replace the second SLM with a camera and slightly adjust the 4-\textit{f} system in front of it with a 400 mm lens, such that the field at the output facet of the RCF is directly imaged onto the camera.
Compared to the earlier experiments, a different camera (Xenics Xeva 320) is used for OAM sorting experiments to avoid the noisy pixel response, which is calibrated for the quantitative analysis of the sorting performance (see Supplementary Note 7). Nevertheless, this camera is not suitable for off-axis digital holography due to the limited number of pixels. In the sorting experiments, we synthesize the input OAM mode and apply the Talbot phase mask (see Supplementary Note 6) simultaneously with the first SLM.

\section*{\textbf{Supplementary Note 5. Ring-core fiber used in the experiments}}
The RCF utilized in the experiments is the same as the one used in a prior publication, wherein the fabrication details are described \cite{eriksson2021talbot}. At 1550 nm, the refractive indices of the ring-core and the cladding of the fiber are around 1.457 and 1.444 \cite{malitson1965interspecimen}, respectively, and their relative refractive index distribution is measured during fabrication using a preform analyzer (Photon Kinetics, P-101). A microscope image of the cross-section of the fiber is shown in Supplementary Fig. \ref{fig_s6}. The RCF pieces of different lengths employed in the experiments are prepared using a conventional fiber cleaver. Limited by the precision of the fiber cleaver used, we slightly tune the laser frequency within C-band to match the exact fraction of the Talbot length of the RCF. Better fiber cleaving techniques or larger core RCF may be used to directly obtain the exact fiber length. 

\begin{figure}
  \renewcommand{\figurename}{Supplementary Figure}
    \centering
    \includegraphics[width=0.75\textwidth]{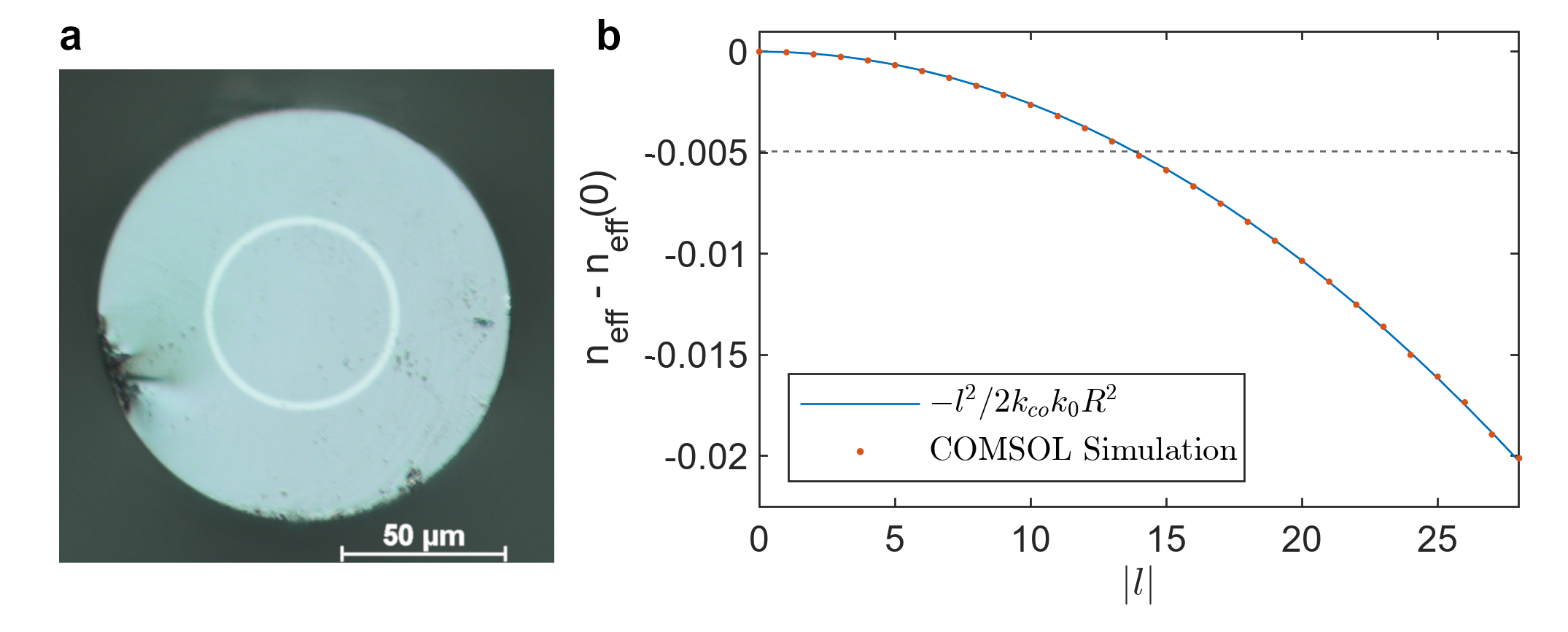}
    \caption{\textbf{a} Optical microscope image of the cross-section of the used RCF. \textbf{b} The OAM dispersion of RCF eigenmodes obtained through theory (blue line) and COMSOL (red dot) simulations. The dashed horizontal line indicates the threshold between bound and leaky modes. Only one effective refractive index is shown from each simulated mode group with equal $|l|$. 
    }
    \label{fig_s6}
\end{figure}

The angular Talbot effect relies on quadratic phase shaping in the OAM degree of freedom, which is naturally realized by light propagation in the RCF. To elaborate, solutions $\psi(\phi, z) = \chi(\varphi, z)\text{e}^{ik_{\text{co}}z}$ to the paraxial Helmholtz equation in an infinitesimally thin cylinder shell of radius $R$ can be written as a Fourier expansion into OAM modes \cite{eriksson2021talbot}:

\begin{equation} \label{eq:s8}
    \chi(\phi, z) = \sum_{l\in \mathbb{Z}} \chi_l\text{e}^{il\phi}\text{exp}\left(-i\frac{zl^2}{2k_{\text{co}}R^2} \right),
\end{equation}

\noindent{where} $\phi, z$ are the cylindrical coordinates, $\chi_l$ are the Fourier amplitudes of the incident field $\chi(\phi,0)$ decomposed in the OAM basis, $l$ is the OAM mode index, and $k_{\text{co}}$ is the wavenumber in the cylinder shell (fiber core). Here, the latter phase factor is a relative phase picked up by the different OAM modes upon propagation along $z$, that is, through the fiber. As is evident from the equation, this propagation-induced phase is quadratically dependent on the OAM mode index $l$, elucidating the origin of the quadratic OAM dispersion found also in RCFs.

To show the validity of this greatly simplified picture, we can compare the effective refractive indices of OAM modes predicted by Eq. \eqref{eq:s8} to those simulated in COMSOL. First, from Eq. \eqref{eq:s8} we derive a general form of OAM dispersion $\beta(l) = k_{\text{co}} - l^2/2k_{\text{co}}R^2$, from which the effective refractive index of each mode is obtained: 
\begin{equation} \label{eq:s9}
    n_{\text{eff}}(l) = \beta(l)/k_0 = n_{\text{co}} - l^2/2k_{\text{co}}k_0R^2,
\end{equation}
\noindent{where} $k_0$ is the vacuum wavenumber, and $n_{\text{co}}$ is the refractive index in the fiber core. The simplifications of the theory result in a constant offset in the effective refractive indices of the modes between simulation and theory, but the quadratic dispersion is accurately modelled. 
To show this, we plot in Supplementary Fig. \ref{fig_s6}b the relative effective refractive indices of OAM modes $n_{\text{eff}}(l) - n_{\text{eff}}(0)$ obtained through COMSOL simulation and from Eq. \eqref{eq:s9} $n_{\text{eff}}(l) - n_{\text{eff}}(0) = - l^2/2k_{\text{co}}k_0R^2$, which are in excellent agreement. The dashed horizontal line in Supplementary Fig. \ref{fig_s6}b indicates the threshold between bound and leaky modes of the RCF. Although the modes below the threshold are intrinsically leaky, they maintain low loss when propagated in the RCF for just a few centimeters as implemented in the experiments. In addition, the leaky modes near the threshold are recently found to create centrifugal barriers by themselves, thus maintain low loss transmission even over kilometer length scales of RCFs \cite{ma2023scaling}. 

\section*{\textbf{Supplementary Note 6. Talbot phase masks used in the experiments}}
\begin{figure}[ht!]
  \renewcommand{\figurename}{Supplementary Figure}
    \centering
    \includegraphics[width=0.75\textwidth]{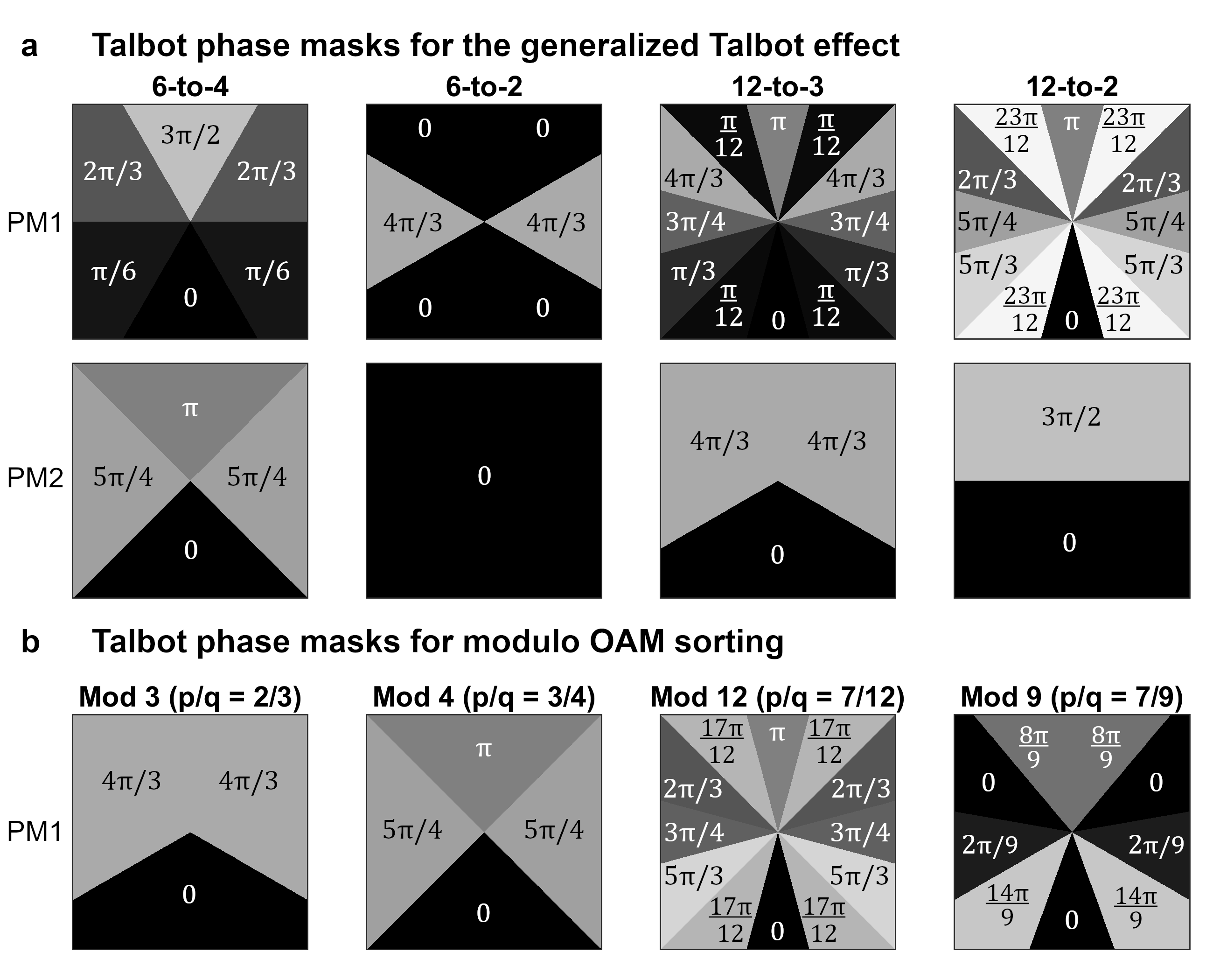}
    \caption{Talbot phase masks used in this work. \textbf{a} Talbot phase masks used for the generalized Talbot effect experiments. $n$-to-$m$ corresponds to the transformation from $n$ in-phase petals to $m$ in-phase petals. PM1 (top) and PM2 (bottom) are the Talbot phase masks applied to SLM 1 and SLM 2, respectively. \textbf{b} Talbot phase masks used for modulo OAM sorting experiments. For a modulo $q$ OAM sorter using $p/q$ of the Talbot length of the RCF, PM1 is the Talbot phase mask applied to SLM 1.}
    \label{fig_s7}
\end{figure}

All Talbot phase masks used in the experiments shown in this paper are pictured in Supplementary Fig. \ref{fig_s7}, excluding the ones already shown in Fig. 2 of the main text. In Supplementary Fig. \ref{fig_s7}a, the top and bottom rows correspond to the Talbot phase masks applied to SLM 1 and SLM 2 in the generalized Talbot effect experiments, respectively. They are calculated from the phases of self-images in a Talbot carpet. Supplementary Fig. \ref{fig_s7}b illustrates the Talbot phase masks applied to SLM 1 in the modulo OAM sorting experiments. These masks are calculated based on Eqs. \eqref{eq1} and \eqref{eq2} in Supplementary Note 1. The $s$ parameter is determined by the length of the RCF used: $s=2$ for $p/q=2/3$, $s=3$ for $p/q =3/4$, $s=7$ for $p/q=7/12$, and $s=4$ for $p/q=7/9$. 

\newpage
\section*{\textbf{Supplementary Note 7. Calibration of the camera}}
In the modulo OAM sorting part of the experiment, the relative power values are calculated by integrating the intensity registered as gray values in the camera, over regions of camera images. For accurate measurements and analyses, the pixel response of the camera must be characterized and linearized. The camera in use (Xenics Xeva 320) hosts a nonlinear pixel response to the intensity of light when uncalibrated, necessitating this procedure. The used characterization procedure consists of first shining a large Gaussian laser beam on the camera, occupying most of the camera sensor. Images of the laser beam are recorded at varying laser power, for a total of 81 images, with laser power ranging from being barely visible on the camera, to oversaturating the central part. This way, the gray values of the recorded images span the entire range of possible values. Simultaneously, the relative power of the laser beam is recorded for each image taken. The dataset can be understood as a 3-dimensional matrix of gray values $G_{\text{raw}}(x,y,p)$, accompanied by the set of recorded power values $P(p)$. $(x, y)$ are the pixel array coordinates in the camera and $p$ is an index corresponding to different laser powers.

The characterization algorithm attempts to find a polynomial function that linearizes the nonlinear pixel response of the camera $G = AG_\text{raw} + BG_\text{raw}^2 + ...$. A cost function is built that iterates through every element in $G(x,y,p)$. For all elements of G($x$,$y$,$p$), the cost function picks a set of gray values from the current camera pixel $(x,y)$ but different power levels $G(x,y,p')$, where $p'\neq p$ is a set of fixed, but randomly chosen power indices for each element in $G(x,y,p)$. The set $p'$ can in principle consist of any number of indices, but for a balance between characterization accuracy and speed we choose 3 power indices $p'$ for each element in $G(x,y,p)$. The ratio of the pixel values at different power levels is compared with the measured ratio of power values: $\sum_{p'} |\frac{G(x,y,p)}{G(x,y,p')}-\frac{P(p)}{P(p')}|$. If the intensity response of the camera is linear, the ratio of gray values is equal to the ratio of the recorded power values, thus the sum is zero. With a nonlinear pixel response, the ratios are in general different, and this sum will yield a nonzero, positive value.
These comparisons span the entire pixel array and range of powers, although neglecting very low ($< 0.1$ of maximum value) and very high ($> 0.95$ of maximum value) gray values to avoid underexposed and oversaturated pixels.

The recorded dataset is inherently biased, containing an unneven distribution of different gray values, in our case containing a higher number of low gray values than high gray values. To take this into account, a histogram of the gray value distribution is recorded, yielding a number of occurrences for each gray value $n(g)$ in the entire 16-bit range of gray values $g \in [0, 65535]$ in the camera. The cost function weights the gray value comparisons by the inverse of the product of the occurrences of both gray values $g_1$ and $g_2$ in each comparison, i.e., by $(n(g_1)n(g_2))^{-1}$. This weighting exactly counteracts the unbalanced number of occurrences between different pixel values. Therefore, the total cost function can be written as $C=\sum_x\sum_y\sum_p\sum_{p'}|\frac{G(x,y,p)}{G(x,y,p')}-\frac{P(p)}{P(p')}|(n(G(x,y,p))n(G(x,y,p')))^{-1}$.

This cost function is minimized by optimizing the coefficients of the linearization polynomial using a built-in Nelder-Mead algorithm in MATLAB. Since in general the nonlinearity of the pixel response can depend on the settings of the camera, the settings are kept fixed across all measurements. Therefore, a single linearization polynomial is valid across all measurements.

\section*{\textbf{Supplementary Note 8. Unconventional modulo 9 OAM sorter ($p=7$, $q=9$)}}
\begin{figure}[h!]
  \renewcommand{\figurename}{Supplementary Figure}
    \centering
    \includegraphics[width=0.95\textwidth]{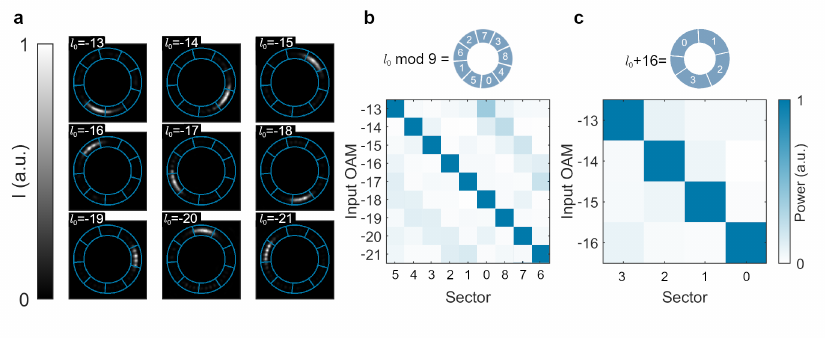}
    \caption{Unconventional modulo 9 OAM mode sorting and enhanced sorting of 4 consecutive OAM modes.  The mode sorters are constructed based on a Talbot phase mask with 9 sectors and an RCF piece with a length of $\frac{7}{9}z_T$. \textbf{a}, Measured intensity distributions at the sorter output for input OAM modes $l_0$ ranging from -21 to -13 (labeled on the top left corners). \textbf{b}, Top: Layout of the sorter for 9 OAM modes. Neighboring OAM modes are mapped to 7 angular sectors apart at the sorter output. Bottom: Crosstalk matrix of the mode sorter, with an average sorting accuracy of $66.6\pm 4.2\%$ degraded by light leakage to neighboring sectors. \textbf{c}, Since any 4 consecutive OAM modes are never imaged to neighboring sectors, a special set of wider sectors can be constructed to sort these modes with a higher accuracy. Top: Layout of the sorter for OAM orders from -16 to -13. Bottom: Crosstalk matrix of the mode sorter, with an average sorting accuracy increased to $84.9\% \pm 2.6\%$.}
    \label{fig_s8}
\end{figure}

In addition to the OAM sorters shown in the main text, we present a modulo 9 OAM sorter constructed from a Talbot phase mask of 9 sectors (see Supplementary Note 6) and an RCF with the length of $\frac{7}{9}z_T \approx 2.33$~cm. Supplementary Fig. \ref{fig_s8}a-b demonstrate the experimental results of the modulo 9 OAM sorter for input OAM modes from -21 to -13, achieving on average $66.6\pm 4.2\%$ of the sorting accuracy. Here, the sorter images every consecutive OAM mode 2 sectors away, which can also be viewed as being 7 sectors away from another direction, thus consistent with the RCF length used in this case. The unconventional sorting behaviour leads to the fact that every 4 consecutive OAM modes are never imaged to neighboring sectors. This can be harnessed to sort 4 consecutive modes with a higher accuracy, by using a set of 4 wider sectors as shown in Supplementary Fig. \ref{fig_s8}c. In this scenario, an improved sorting accuracy of $84.9\% \pm 2.6\%$ is experimentally achieved. Similarly as done in the main text, considering the central $90\%$ and $75\%$ of all sector areas yield further improved sorting accuracies of $87.8\pm 1.2\%$ and $86.1\pm 2.1\%$, respectively.

\newpage
\renewcommand{\bibpreamble}{
$^\dagger$These authors contributed equally to this work.\\
$^\ast${Corresponding author: \textcolor{magenta}{jianqi.hu@lkb.ens.fr}}\\
}

\bibliographystyle{naturemag}
\bibliography{ref}